# A gravitating pattern unifies the living and physical worlds

Liz Wirtanen

Kativik Ilisarniliriniq, Salluit, Canada

August 15, 2021

**The inflationary hot big bang model of cosmology explains the birth of the universe, the appearance of atoms and that of large-scale structures, altogether omitting any reference to the living world. Yet, like the physical world, the living world occupies space, is composed of baryonic matter, and is submitted to the four fundamental forces. To bridge this break in continuity, a study relying solely on the laws of physics was conducted to seek a pattern that would unite both worlds. Here we show evidence of a common pattern that is based on a gravitating organization. Seven criteria were worked out to define a gravitating concept that applies to any attractive force. A scan through the second version of the Integrative levels classification system revealed that the following six types of objects satisfy the criteria: atoms, eukaryotic cells, vertebrates, nation-states, planetary systems, and galaxies. A comparative diameter study was carried out on objects confined to the Solar System in the Milky Way. The trendline linking the average diameters follows an exponential function $f(x)=0.2527 \times 10^{6.23x}$, with a coefficient of determination of 0.9904. The objects assemble into a superstructure spanning 32 orders of magnitude, systematically skipping 6.23 orders of magnitude at each level. The unity of structure revealed in this study suggests that both worlds result from a common underlying organizing process. Furthermore, the gradual and uninterrupted sequence leading to complex structures supports the idea that the gravitating organization offers a favorable pathway to complexity. The framework presented in this study may help address complex phenomena such as emergence and agency, broaden the scope of the theory of evolution and fuel the debate on the definition of life itself. Ultimately, it may contribute to a more complete understanding of the universe.**

**Introduction**
Cosmology brings together astronomers and physicists in a joint effort to understand the physical universe as a unified whole[1,2,3]. The life sciences are not included in this exercise, but as life is also a part of the universe, its study may have relevance to understanding the whole. To this day, no model integrates the theory of evolution[4] with cosmology. However, innovative approaches drawing upon information theory are providing insights that may redefine cosmology[5] and join the living and physical worlds in a unitary framework[6]. To pursue the process, a study was conducted to find a pattern linking both worlds.

Entities are currently classified in a system derived from Feibleman's observation that the natural world is organized in a series of levels of increasing complexity, that each level cannot exist without the lower ones and that each level displays emergent properties not found at lower levels[7]. The most recent Integrative Level Classification system (ILC 2) comprises six strata (information, matter, life, mind, society, and culture) inspired from Hartmann[8]. Divided into 10 851 classes, its database provides an exhaustive list of all known phenomena in the universe[9,10]. The classification system, however, lacks a formal framework based on the laws of physics, that would explain how the levels of complexity become established.

Hartmann considers unity of structure, or an all-pervasive order, as being reconcilable with the far-reaching heterogeneity of the world. The starting point to explore this idea was to look at how a planetary system self-assembles. A gravitating organization was assumed to be the most probable pattern unifying the living and physical worlds because it is common in the physical world, three fundamental forces (strong nuclear, opposite charge attraction, and gravity) oppose the expansion of the universe, and the nested integrative levels of the natural world suggest a stepwise, repeated inward collapse of matter.

The gravitational organization of large-scale structures is well documented[11], but little attention has been given to other objects. A planetary model of the atom was proposed in 1911 after Rutherford's discovery of the atomic nucleus, but Schrödinger's probabilistic model eventually prevailed[12,13]. The idea of a gravitational organization was discarded on the basis that opposite charge attraction was the force holding the atom together, rather than gravity. However, according to the Oxford dictionary, to gravitate is *to move, or tend to move, toward a center of gravity or other attractive force*[14]. Gravity may not be the only force that holds objects in the vicinity of a center. The attraction and maintenance of electrons in stable probabilistic orbits near the atomic nucleus argues for a gravitational-like organization. The fact that this type of organization was displayed at both the large and small scales suggested it could also occur in the middle range, where life appeared.

**Methods**
No clear definition of a gravitational system exists. A series of criteria was therefore drawn up to help define the concept and extend it to any attractive force. Accordingly, an object is considered a gravitating system if it satisfies the seven following criteria:

1. It is composed of a distinctive core surrounded by peripheral elements.
2. Its core predominates over the peripheral elements.
3. It has a size limit.
4. It presents mechanisms preventing inward and outward collapse.
5. It goes through an assembly process, has a lifespan, and ends up destroyed.
6. Its constituents can be recycled.
7. It displays emergent structuring capacities.

The term *gravitating* is introduced to avoid confusion with the word *gravitational*. The set of criteria was deduced from the organization of the Solar System. It has a central star whose distinctive feature is nuclear fusion reactions. It is surrounded by peripheral elements. The planets are bound to the sun by gravity. The system has a size limit beyond which gravity has no binding power. Inward collapse is avoided by nuclear fusion reactions and by planetary motion. Outward collapse is prevented by gravity. The Solar System is assembled from interstellar gases and dust, has a lifespan, will eventually be destroyed, and its parts may be recycled. Along with other similar systems, it organizes into a galaxy. Planet-satellite assemblies are considered subsystems because of their hierarchical relationship with the core. Objects that do not fit the pattern, such as molecules, prokaryotic cells, and multicellular plants, are considered intermediate forms or subsystems.

The scan for the gravitating pattern was conducted in the ILC 2 system. The Avg-Max-Min Chart was plotted using the Microsoft Excel program. The comparative diameter study includes the largest possible size ranges of atoms, eukaryotic cells, vertebrates, and nation-states. Diameters of atoms, and eukaryotic cells were calculated assuming a spherical shape. Vertebrate length was used to approximate diameter. Diameters of nation-states were calculated assuming a flat circular shape. For single objects such as the Solar System and Milky Way, maximal and minimal diameters based on available boundary values were used. The study focuses on the Solar System and Milky Way, as they are the only large-scale structures known to host life. Finally, many planets have been detected in recent years, supporting the view that the planetary organization is the norm rather than the exception[15]. For the remainder for the study, it will be assumed to be the case.

**Results**
**The gravitating pattern**
Based on the seven criteria presented in the Methods section, a systematic scan through the ILC 2 system revealed that the following types of objects display a gravitating organization: atoms, eukaryotic cells, vertebrates, nation-states, planetary systems, and galaxies. The characteristics of the objects retained in the study are presented in table 1, by order of increasing diameter. Below, they are described according to their order of appearance in the universe.

The first gravitating system to make its appearance in the universe is the atom. The atom has a positively charged core attracting a series of negatively charged peripheral electrons. Electrons farthest from the nucleus are least attracted while the deeper levels are held in place more securely. Beyond a certain size, the structure collapses, setting an upper limit to atomic size. Proton blowout is avoided by the presence of neutrons and crashing of electrons into the nucleus is avoided by their kinetic energy. Atoms go through a complex series of assembly steps, beginning with nucleosynthesis, and ending with the acquisition of wandering electrons in space. Atoms have the longest lifespan, extending beyond the theoretical lifespan of the universe. They demonstrate two distinct structuring properties, one via their electron structure that allows atoms to assemble into molecules through opposite charge attraction, and the second, by the action of bulk mass modifying space. Molecules extend the variety of building blocks available for the next steps.

| Details | Objects | | | | | |
|---|---|---|---|---|---|---|
| | Atoms | Eukaryotic cells | Vertebrates | Nation-states | Solar System | Milky Way |
| **Core** | Nucleus | Nucleus | Heart | Parliament | Star | Black hole |
| **Core distinctive feature** | Positive charge | Genetic information | Pump | Government | Nuclear fusion reactions | Singularity |
| **Major peripheral elements** | Electrons | Organelles | Organs | Citizens | Planets | Stars |
| **Binding agent** | EM attraction | EM attraction | EM attraction | Information | Gravity | Gravity |
| **Size ranges (raw data)** | 31 to 244 pm (radius)[16] | 10 to 100 μm (diameter)[17] | 7.7 mm to 33 m (length)[18,19] | 2 to 17 098 242 km$^2$ (area)[20] | 9.09 x 10$^9$ km to 1921.56 AU (diameter)[21] | 25 to 290 kpc (radius)[22,23] |
| **Diameter ranges** | 62 to 488 pm | 10 to 100 μm | 0.0077 to 33 m | 1.6 to 4666 km | 60.763 to 1921.56 AU | .16 to 1.9 MLY |
| **Average diameter** | 275 pm | 55 μm | 16.5 m | 2333 km | 991.16 AU | 1.03 MLY |
| **Average diameter in Angstroms** | 2.75 x 10$^0$ | 5.5 x 10$^5$ | 1.65 x 10$^{11}$ | 2.33 x 10$^{16}$ | 1.48 x 10$^{24}$ | 9.74 x 10$^{31}$ |
| **Size limited in part by:** | Nuclear drip line[24] | Surface to volume ratio[17] | Scaling interactions[25] | Politico-economic forces[26] | Stellar mass[27] | Radiation vs dark matter density, possibly[28] |
| **Inward collapse prevented in part by:** | Electron kinetic energy | EM repulsion, cytoskeleton | EM repulsion, skeleton | Minimal space requirement | Nuclear fusion reactions, planetary revolution | Star revolution, possibly dark matter[29] |
| **Outward collapse prevented in part by:** | Strong nuclear force, neutrons, EM attraction | EM attraction, cytoskeleton | EM attraction, skeleton | Laws | Gravity | Gravity, possibly dark matter[29] |
| **Lifespan ranges** | Over 10$^{32}$ years[3] | Days to years[30] | Months to years[31] | Centuries[32] | Gigayears[33] | Undetermined |
| **Approx. number of objects per level** | – | 10$^{14}$ atoms[34*] | 10$^{14}$ eukaryotic cells[34*] | 3 x 10$^7$ citizens[35] | 195 nation-states[36] / > 10$^{54}$ atoms[37] | 10$^{11}$ Stars[38] |

**Table 1.** Main characteristics of the objects displaying a gravitating pattern. Abbreviations: EM - electromagnetic; AU - Astronomical Units; MLY - million light years; Approx. - approximate. * In the human body.

The Solar System results from the inward collapsing of large quantities of bulk matter under the action of gravity. Its central star is a dense core of plasma around which planets gravitate. Like in the case of electrons, the most distant objects are most loosely connected to the whole. They maintain their orbit at a slower pace, while closer to the core, planets are moving more rapidly. The planetary system has a size range beyond which gravity does not act, as gravity quickly diminishes according to distance squared. Within the stars, nuclear fusion reactions balance the inward pull of gravity, maintaining a stable radius. Planetary revolution prevents collapse of the periphery into the core. Planetary systems develop, have a lifespan, and die in a spectacular explosion of their central star, spreading new types of materials into space. Some of this material gets recycled into other planetary systems. They demonstrate a basic structuring capacity by assembling into galaxies. On a larger scale, galaxies form from the inward collapsing of stars under the action of gravity. The Milky Way has a dense core, a supermassive black hole, that contributes to holding the galaxy together. Dark matter haloes are thought to be producing extra gravity within galaxies. Their presence forces the outer stars to revolve at constant velocity rather than a gradually slower motion expected according to distance from center[39]. Like other gravitating systems, galaxies have a size range, a lifespan and organize into higher-level structures.

Eukaryotic cells are composed of molecules held in place by the action of opposite charge attraction. They have a core, the nucleus, and a periphery, consisting of a set of organelles located in the cytoplasm. Nucleus transfer experiments demonstrate the predominance of the core over the periphery[40]. Cells have a defined size range limited by surface to volume ratio. A rigid structure, the cytoskeleton, keeps the cytoplasm from collapsing into the nucleus, and preserves the structural and functional integrity of the cell. Cells go through an assembly process, have a relatively short lifespan, and die, enriching their environment with organic matter. This matter in turn, will be recycled into other living organisms. Many emergent properties appear with cells, one of which is the surprising capacity to make more of the same. Through cell division and growth, resources are assimilated and assembled into replicas of the original. The DNA contained in the nucleus holds a linear sequence of information which directs biochemical processes in the cytoplasm, speeding up assembly by avoiding molecules to randomly meet. The transmission of genetic information from one generation to the next ensures continuity through time of this remarkable structuring capacity. Through self-replication of the first prokaryotic cells, the abiotic world became the living. Eukaryotic cells also demonstrate other structuring properties such as a capacity to assemble into complex multicellular organisms.

Multicellular organisms such as vertebrates, are communities of eukaryotic cells held in place by molecules that form bonds at cell junctions. The human body will serve to illustrate this complex adaptive system. It is composed of a core, the heart, surrounded by peripheral organs. The heart is required for the periphery to survive; it connects with all cells of the body, tending to each one's needs via the blood. Close to the heart, blood speed is maximal, but as it reaches the periphery, it slows down to the point of near standstill to allow exchange of gases, nutrients, and waste with individual cells. To visualize this gravitational-like motion, imagine the body to be transparent as water or glass. What appears inside is a stream of particles flowing from the heart to peripheral organs, and back, accelerating close to the heart and slowing down at the outer limits of the system. The organism has a defined size range which is controlled by many factors including surface to volume ratio and growth hormones. Functional integrity is maintained by homeostasis and structural integrity is ensured by the skeleton. The organism goes through an assembly process, has a medium range lifespan, and dies, enriching its environment with organic matter. Some of this matter in turn, will be recycled into other living organisms. Humans engage in numerous structuring activities such as raising families, inventing technologies, writing articles, or producing art. This emergent structuring capacity is commonly referred to as living with purpose, or agency[41]. Humans are built from roughly 29 different elements present in the Earth's crust, but human behavior is hugely different from that of a rock. The connected atoms form an open system absorbing energy and using it to structure the physical and living worlds with unprecedented efficiency. Humans demonstrate their structuring capacity in countless ways, one of them being the building of nation-states.

The final object of the study is the self-determined nation-state. Humans gravitate around strong leaders who establish governments in central quarters, forming a core that rules human activities. The parliament binds all citizens of a given country under its constitution, with laws that include management of all resources on their land. The binding agent, information, safeguards and propagates culture, knowledge, and technological advances. Each country has a fixed size, frontiers are defended by armies. Inward collapse is prevented by minimal space required for individual survival, while societal collapse is prevented in part by laws. Civilizations appear and crumble. What about structuring properties? Nation-states provide the conditions required to attain a collective-level structuring capacity. Great collective accomplishments include space exploration, megastructures, vaccines, particle accelerators, and telecommunications.

**Comparing diameters**
As a first step toward characterizing the gravitating pattern, diameters of the systems were compared using an Avg-Max-Min Chart (See Fig. 1). The plot reveals a trendline connecting average diameters according to an exponential function $f(x)=0.2527 \times 10^{6.23x}$ (where x is an integer referring to a level), with a coefficient of determination of 0.9904. This means the diameter of each type of gravitating system is systematically larger than the preceding one by 6.23 orders of magnitude, or by a factor of 2 million (double of the radius). The high $R^2$ score indicates a tight agreement between the average diameters and the regression model. Additionally, the diameter of each type of object is relatively constant. Sizes are confined within one order of magnitude for atoms and eukaryotic cells, within four orders of magnitude for vertebrates and within three orders of magnitude for nation-states. Size ranges are unavailable for planetary systems and galaxies as the study focuses on the Solar System and Milky Way.

The following example will help grasp the sense of size of the objects compared in figure 1. Six orders of magnitude correspond to a 3 mm flea jumping around in a 300-meter Olympic stadium. This proportion is conserved, climbing the size scale in five giant steps, from atoms to eukaryotic cells, eukaryotic cells to vertebrates, vertebrates to nation-states, nation-states to the Solar System, and the Solar System to the Milky Way. This breadth is difficult to imagine, even more so considering atoms are mostly empty space. The nucleus of an atom is comparable to a 3 cm roach in the stadium (the nucleus is $10^{-5}$ the diameter of the atom[42]). Given these overwhelming dimensions, size variability within groups seems insignificant. A variation one order of magnitude would extend from a 3 mm flea to a 3 cm roach. A variation of one order of magnitude above and below average would span from a .3 mm dust mite to a 3 cm roach, and a variation of two orders of magnitude above and below average would span from a 30-micron amoeba to a 30 cm squirrel. It should be noted that the study is preliminary and includes the greatest possible number of objects. Further studies with more stringent selection criteria may result in smaller size ranges within classes of objects.

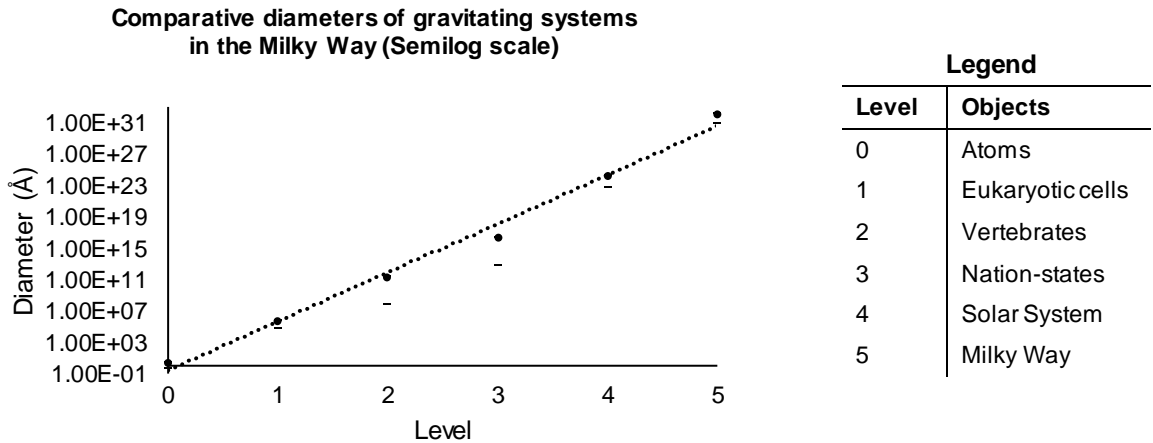

**Figure 1.** Comparative diameters of gravitating systems in the Milky Way. Data is from Table 1. Maxima and minima are represented with horizontal lines. Levels are identified in the legend.

**The superstructure**
The systems form a 6-level superstructure spanning 32 orders of magnitude, illustrated in figure 2. The levels are presented by order of increasing diameter, the smaller ones nested in the larger ones. Each level results from the assembly of elements of the level directly below it, except for the Solar System which skips 3 levels (1, 2, and 3). Atoms either follow a pathway into the living world or skip 3 levels to produce large-scale structures. The large-scale structures (levels 4 and 5) are established first, followed by the living, if the Goldilocks conditions are met[43]. These conditions define the habitable zone around a star, at a specific distance from its center, at a favorable distance from the center of a galaxy. Each planetary system most likely has a habitable zone, whether occupied or not by the living.

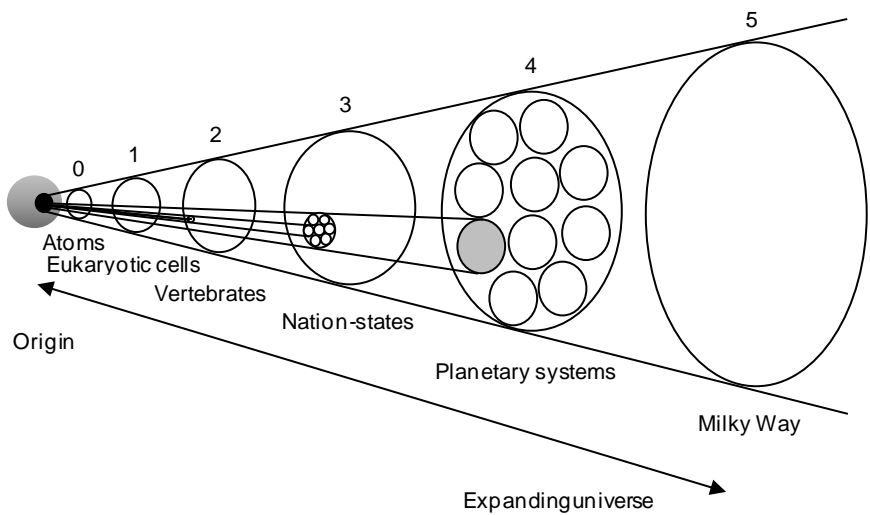

**Figure 2.** Stretched view of the gravitating systems in the Milky Way. Levels are indicated on top. Systems are indicated on the bottom. The Solar System is colored in grey. The origin of the universe is represented by a sphere.

Individual objects grow to a specific diameter, and not beyond. Accumulation of objects within a level increases the diameter of that level until a new organization emerges. There are approximately $10^{14}$ atoms in human eukaryotic cells and $10^{14}$ eukaryotic cells in the human body (See Table 1). An average of $3 \times 10^7$ citizens populates a typical nation-state, and 195 nation-states are confined to the surface of Earth. Over $10^{54}$ atoms make up the Solar System, approximately $10^{11}$ stars are distributed along the spiralling arms of the Milky Way and $10^{11}$ galaxies form the cosmic web of the observable universe[44].

## Discussion
### A world of gravitating systems
At first look, it may seem unreasonable to view the objects presented in this study as gravitating systems. However, if one carefully defines what this type of organization should look like, he or she would come to conclude that the standard gravity-bound system is but one of many types of objects that display a gravitating organization. All systems described in this study have peripheral parts organized around a core. In the physical world, the core provides the attractive force that hold the parts together, while kinetic energy pushes the parts away from the core, at a safe distance. In eukaryotic cells, information shuttles into the core in the form of signalling molecules, then out of it in the form of messenger RNAs, to modify cellular structures and functions. In vertebrates, molecules circulate from the heart to peripheral organs, and back, to ensure survival of the whole structure. Finally, in the nation-state, information travels back and forth from centralized databases to help manage activities of the collectivity. The gravitating pattern is characterised by opposing movements centered on a core. This widespread consistency of form supports the idea that the universe is built on a unitary framework.

The streamlined size distribution illustrated in figure 1, with an $R^2$ above 99%, stands out as dumbfounding. Also surprising is the observation that each type of object has a relatively constant size. There are no atoms the size of a human, no humans the size of a cell, no planetary systems the size of a galaxy, and no galaxies the size of a nation-state. Size limits can be imposed by force ranges, but to account for the superstructure illustrated in figure 2, some underlying mechanism must ensure that these ranges will be complementary and non overlapping. Competition for resources and space within a level may also help limit the size of adjacent structures.

Presumptive levels of organization may be extrapolated from both extremes of the graph. Four possible levels extend below the atom, at $10^{-6}$ Å, $10^{-12}$ Å, $10^{-19}$ Å, and $10^{-25}$ Å (Planck length). Above the galaxy, an extra level at $10^{37}$ Å corresponds to the diameter of the observable universe. Whether organizations lurk at these levels is open to speculation.

### A niche for the living
Figure 3 shows the separation of forces in the very early universe[45]. The strong nuclear force acts exclusively within the short range of $10^{-15}$ m, while gravity produces effects at large to infinite distances[46]. The infinite range of electromagnetism lies in between the two others, filling the gap. The range specificity of these forces reveals an important cleavage concerning the objects they influence. Short range nuclear forces secure atomic nuclei; short to long-range electromagnetic forces hold atoms, molecules, eukaryotic cells, vertebrates, and nation-states; and long-range gravity binds planetary systems and galaxies.

Forces produce their effects on specific targets: strong nuclear force influences quarks, gravity acts on bulk mass and electromagnetism affects electrons. This ultimately determines the constituents of the gravitating systems. Large-scale structures, requiring only bulk matter, can form by accumulation of simple constituents, making no distinction between the various of elements of the periodic table. The living world, on the other hand, exploits the full spectrum of elements of the periodic table, benefiting from the broad associative potential of electrons.

As illustrated in figure 2, space is equally divided between both worlds: 3 levels are occupied by the physical world (atoms, planetary systems, and galaxies) and 3 by the living world (eukaryotic cells, vertebrates, and nation-states). The living world bridges the gap between the small scale and the large scale by filling the medium range, thereby achieving unity of the superstructure.

The marked division of forces, materials, and space provides a fundamental basis to distinguish the physical and living worlds.

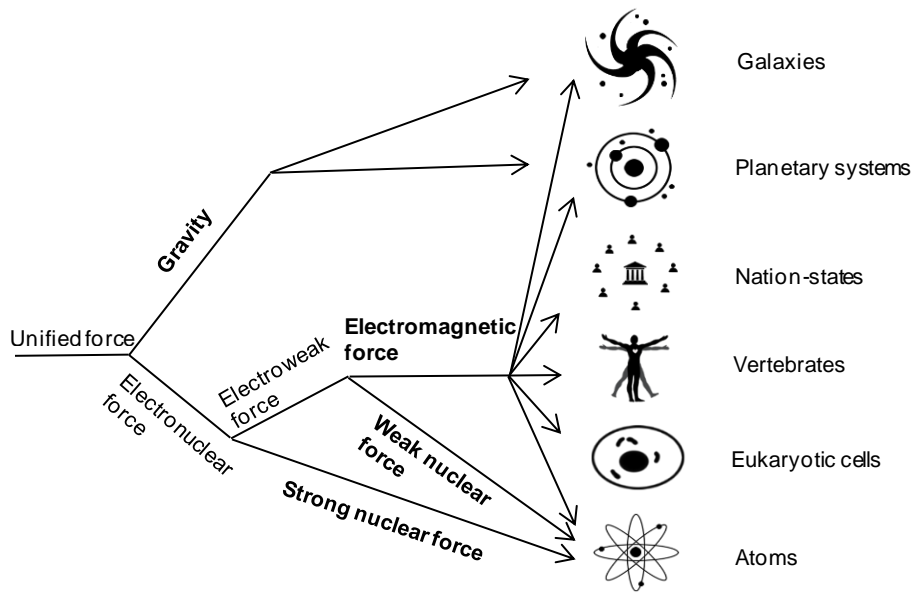

**Figure 3.** Separation of the fundamental forces in the very early universe and ranges of action.

Based on the observations: 1) that the living and physical worlds share a similar gravitating organization; 2) that fundamental forces show ranges specific to each world; 3) that atoms display properties that specifically satisfy the requirements of each world; and 4) that space is evenly divided between both worlds; we posit that the living world is as much a part of the universe as is the physical world. The balanced partition of forces, materials and space between both worlds puts them on equal footing, each contributing relevant objects to the universe. Furthermore, the shared gravitating pattern supports the view that the living world is part of a universal iterative process that began with the big bang, produced atoms, large-scales structures, living organisms and finally, nation-states. This body of evidence strongly argues against the view that life is a fortuitous event resulting from a set of random circumstances. One world cannot be considered a fluke without including the other. Lastly, the secured domain augments the probability of life flourishing elsewhere.

Legitimacy being established, it is now possible to examine the contribution of the living to the whole. The next sections take a closer look at agency, the theory of evolution, and increasing complexity.

**Decision making at the service of order**
Assembly of the gravitating systems is accomplished by way of emergent structuring properties. Each object has its set of capacities: atoms use charges, cells rely on genetic information, citizens vote, bulk mass modifies space. Decision making in complex adaptive systems is an emergent property that can certainly facilitate assembly of the most complex systems. Recent studies in information theory are providing evidence that this unique ability may be supported by the uncertainty principle of the quantum world [6,47].

Decision making requires analyzing multiple courses of action, selecting the best one and imposing it to elements under one's control, to achieve a specific goal. This type of activity is not possible in a deterministic system in which outcomes can be predicted from initial conditions. At the other extreme, decision making cannot be based on randomness because it means no decision is being made. Between these limits, the quantum world supports what seems to be intentional choice. According to Turing, information flowing in a computer can produce unpredictable outcomes [48]. This suggests a degree of independence of electrons from their material support. More recently, Walker's team analyzed the control circuit that regulates the expression of ten genes involved in the cell cycle of fission yeast, a simple unicellular eukaryote [49]. Devising a computer simulation of the network and comparing it to reality revealed that the flow of information in the biological network was well above random. Four out of the ten genes acted as a control kernel, correcting mistakes, and steering the network toward successful cell fission. In addition, the flow of information did not connect regulatory proteins in a causal sequence, suggesting that information flow and processing was happening at another level. Davies, in his most recent book *The Demon in the Machine*, questions whether non-trivial quantum effects such as tunnelling, could be at play in biology [6]. The capacity to analyse information and to change a course of action is obvious in humans, but these studies extend the ability to simple unicellular life forms, bypassing the need for neuronal networks.

**Extending the theory of evolution**
The unity of structure described earlier invites to reconsider evolution from a broader perspective. In his masterpiece *On the Origin of Species*, Darwin defines natural selection as the process by which organisms change through time as a result of changes in heritable physical or behavioral traits. He summarized the concept as *descent with modification*[4]. The process depends on the genetic code. While reproduction is happening, random point mutations and other chromosomal rearrangements are transmitted from parents to offspring. Organisms with novel traits best adapted to their environment are better competitors; they are most likely to win the fight for survival and produce offspring with the same traits. By this mechanism, useful traits spread to entire populations and new species eventually appear.

Excluding genetic transmission, the general principles of evolution seem to apply to all gravitating systems. *Survival* means to maintain structural and functional integrity for a period, in an environment. All systems described in this study survive in their environment for a given time. *Natural selection* favors organisms that possess traits best adapted to their environment. The ubiquity of the gravitating organization suggests it is selected as the most suitable form in any environment. *Competition* for resources and space is an important key to survival. All structures in this study result from a competition for resources and space. In the living, reproduction involves transmission of genetic information. In all gravitating systems, the same pattern is reproduced, over and over again, with no apparent transmission of information. Finally, *change through time* refers to changes happening from one generation to the next. All objects described in this study change through time, resulting in a diversity of atoms, eukaryotic cells, vertebrates, nation-states, planetary systems, and galaxies.

From the above, it appears that the concept of evolution can be broadened. Everywhere in the universe, transient structures appear and dissolve, giving rise to a variety of objects. The physical world, like the living one, seems to be evolving. This should not come as a surprise because the living and the physical worlds involve the same natural elements submitted to the same fundamental forces. The present study shows that diversification happens along two different axes: within a level, and along the size scale.

**A favorable pathway to complexity**
Diversification is associated with complexity, the most prominent trait of the living world. In this study, a complex system is defined as *a system composed of many components which may interact with each other*[50]. Complexity is considered to increases as new systems appear in the universe because their components can interact with each other as well as with components of all preceding levels. For example, chlorophyll in living organisms interacts with electromagnetic radiation originating from a large-scale structure established before the living world. The gradual and uninterrupted road to complexity articulated around a gravitating pattern supports the idea that this organization offers a favorable pathway to complexity. Life forms such as multicellular plants and lower animals, based on other patterns, do not achieve higher-level organizations.

Complexity also applies to emergence. According to systems theory, *emergence occurs when an entity is observed to have properties its parts do not have on their own, properties or behaviors which emerge only when the parts interact in a wider whole*[51]. As gravitating systems gain complexity, they present graded emergent structuring capacities. Bulk mass generates the limited structuring capacity of large-scale structures; lightweight electrons support the spectacular behaviors of the living world; and the ethereal photon, carrying no mass or charge, propels nation-states forward by transmitting digital information at the speed of light. Emergence gains complexity and freedom as material support wanes.

The trend seems to extend to abstract forms, which bear no mass, charge, or energy. An equation such as Einstein's $E=mc^2$ has impacted the world significantly despite its trivial structure. Mathematical relations exist regardless of any support, as an abstraction of reality. They exist regardless of the existence of the universe itself, as a possibility. Abstract information is indestructible and can be concentrated infinitely everywhere. Were the universe to collapse inward back to nothingness, it should concentrate to a point of bodiless infinite possibilities, the ultimate abstract form. Yet, in the absence of any known support, the origin has expanded the universe into existence, providing an example of an abstract form displaying emergence.

The organizations described above can be classified into the four groups illustrated in figure 4. The trend suggests that abstract forms have the greatest potential for complexity and freedom. Creation of a universe from nothing may represent but one of several capacities of nothingness. Moreover, the line of reasoning fuels the debate on the definition of life itself, by putting forward the provocative idea that nothingness may be an abstract form of life.

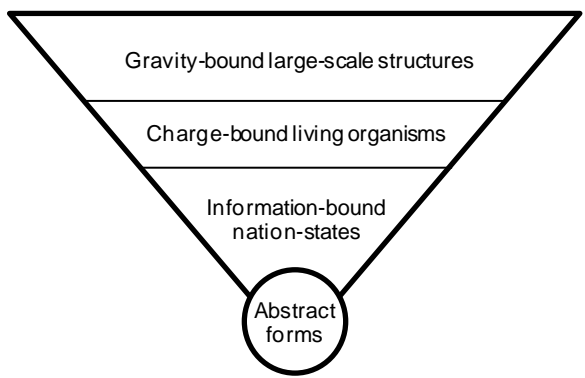

**Figure 4.** Classes of forms.

The origin of the universe and the universe per se, considered together, seem to adopt the gravitating pattern. Figure 4 can be viewed in three dimensions as a set of four concentric spheres. At the center, nothingness defines an abstract core, and is surrounded by material structures that burst out of it. The core is distinctive by its abstract nature and takes precedence over the universe because it caused its existence. The universe tends to collapse inward toward the core, at least locally, but the inward collapse is prevented by expansion. The universe goes through an assembly process, most likely has a lifespan and possibly a size limitation. Last but not least, the core may be the ultimate source of emergence. In short, the universe, like other gravitating systems, seems to be characterised by opposing movements centered on a core.

Because of expansion, the core may be dispersed everywhere in space, most stretched in the intergalactic space and least stretched in quarks. However, on account of its abstract nature, it could remain whole regardless of stretching, supporting Bohm's belief that the whole is in all the parts. In *Wholeness and the Implicate Order*, he describes a total order enfolded in each region of space and time[52]. He compares the universe to a hologram in which each part of reality contains information about the whole. The hectic quantum activity inherent to the vacuum, or empty space[3], provides physical evidence that nothingness may be active everywhere.

Two key concepts brought to light in this study, unity and freedom, can drastically remodel a world view. To conclude the discussion, a third potentially ground-breaking notion is explored: the idea of an organizing principle behind the evolution of the universe.

**A framework based on opposing forces**
Each level of the superstructure illustrated in figure 2 results from the interaction of two opposing forces. Expansion, originating in the big bang, pushes bits and pieces away from one another, increasing disorder, or entropy. Attractive forces, on the other hand, oppose the expansion by pulling parts inward toward the origin, installing order, reversing entropy. The inward collapse brings pieces closer together, shortening the diameter separating them, closing in to a zero diameter. Each object of the superstructure, however, does not completely collapse inward to oblivion. Rather, peripheral elements are locked in a position at a safe distance from the core, trapped in a tug of war between inward collapse and outward tearing. Based on these general considerations, the following protocol for building the universe is proposed:

> *The outward expansion of the universe leads to the separation of the unified force into four fundamental forces, while the gradual stepwise inward collapse allows those forces to organize matter, energy, and information according to a basic gravitating pattern, from simple to complex, with some overlapping between levels and some intermediate steps.*

The gravitating systems presented in this article all defy entropy for the duration of their life. They exist because their connected pieces withstand the tearing force of expansion. If no force opposed entropy, all pieces would simply fly away from one another as the universe expanded. Atoms would explode; living organisms and nation-states would collapse; no planetary systems or galaxies would ever appear. To gain insight on this complex interplay of opposing forces, consider nothingness to be a highly stable ground state. An important quantity of energy such as a big bang is required to stretch it out. A stable structure naturally resists stretching, by contracting back like an elastic fabric, giving rise to apparent attractive forces. Waddington's epigenetic landscape metaphor can serve to illustrate this spontaneous return to stability[40]. The balls in figure 5 can represent atoms. Placing them at the top of the hill requires energy but their subsequent rolling down the slope is spontaneous. They naturally tend towards the lowest energy level. Their random downward paths lead to the various structures of the universe. Attractive forces invariably produce organized structures because order is more stable than disorder, albeit temporarily.

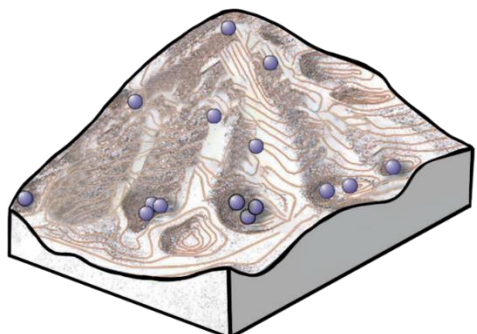

**Figure 5.** Waddington's landscape model.

**A grid to facilitate assembly**
Opposing forces, because they tend to cancel one another, are not sufficient to establish gravitating systems. Some asymmetry is required to avoid annihilation. A grid may help the assembly process as well. Many grids and frameworks that act at specific levels in the universe have been described. At the quantum level, the Higgs scalar field provides a numerical network leading to differentiation of fundamental particles[3]. Random quantum density peaks appearing in the very early universe are the site of accumulation of dark matter much later, responsible for producing the well structured cosmic filamentous web[53,54]. At the atomic level, Jedamzik's group recently did computer simulations in which weak magnetic fields were added to a simulated plasma-filled young universe[55]. Protons and electrons in the plasma followed the magnetic field lines and accumulated in regions of weakest field strength. This convergence facilitated the assembly of hydrogen atoms. The living world provides examples of 3-dimensional grids involved in metazoan development[56]. Transcription factors responding differentially to morphogen gradients activate location-specific genes during development, resulting in cell differentiation and proper positioning of organs in the organism. The process explains how the same genetic code, despite being present in each cell, is expressed differentially, according to position in the organism. It also provides an example of an abstract, enfolded, complete version of the structure, lying in each part.

A grid at the scale of the universe might ensure proper alignment of levels relative to one another other, like a skeleton ensures proper positioning of organs in an organism, essential for functional integrity. It should demonstrate a gradation from abstract to bulky and unfold as the universe expands, prior to assembly. Magnetic fields, because they permeate the universe, could be good candidates for such a structuring grid[57]. Other contenders include high energy waves, energy gradients, electric fields, dark energy, dark matter, the shape of the universe, and ordinary matter.

**How to build a universe in a single shot**
According to Guth, all the mass of the universe may have evolved from an initial seed smaller than a proton. Inflation would have expanded the seed to a radius of 1 meter, followed by the regular rate of expansion of the big bang[3]. Because this scenario involves the inherent randomness of the quantum field, an infinite set of universes, each with different laws of physics, may have bubbled from nothingness, producing universes that failed and others that worked. A more efficient approach to building a functional universe is advanced here. In this scenario, processing of information and decision making could be carried out by nothingness, the ultimate organizing principle. It has been mentioned previously that these processes may be taking place at levels more fundamental than expected. Additionally, in accordance with the scheme of this study, nothingness may predominate over the universe and have other capacities besides creation of matter and energy. The alternative scenario will be discussed using the example of a terrestrial construction project.

A major construction project starts with an abstract idea, followed by sketches and technical drawings. The construction phase requires guiding tools like strings or lasers to produce regular geometric shapes, formworks for foundations, and scaffoldings to ensure proper construction of walls. Each of these structuring tools contains information on some type of support. They start in the abstract realm of infinite possibilities, continue in 2 dimensions on a piece of paper and finish in 3 dimensions, in the physical word. They are not part of the final object but were present on a temporary basis to ensure proper assembly of parts. Most importantly, they precede the work. The progression from an initial idea to a well constructed functional universe could follow the same logical sequence as above. An abstract structure may have processed the entire universe project before execution, weighing each option before making a choice, weeding out failures. It may have programmed the space, energy, forces, materials, grids, and frameworks required at each step, in essence, determining the laws of physics. The autonomous execution that followed could have led to the variety of gravitating systems described in this article. If this model holds, our universe may turn out to be a confined subset of a complex abstract structure whose properties are not yet fully understood. Unveiling information processing mechanisms at the most fundamental level will prove essential to substantiate this conjecture.

**Concluding remarks**

The word universe literally means *turned into one*. In this study, we report the characterization of a pattern that links the living and physical worlds. We show that objects of both worlds adopt a similar gravitating organization and assemble into a superstructure spanning 32 orders of magnitude. The unity of structure suggests that both worlds result from a common underlying organizing process. Furthermore, the gradual and uninterrupted sequence leading to complex structures supports the idea that the gravitating organization offers a favorable pathway to complexity. The unitary framework that emerges from this study constitutes a rough draft that may serve as the premise to develop a more complete understanding of the universe. Quantifying complexity[58], studying matter and energy distributions inside each level, and how the levels interconnect, may lead to a new area of investigation. The model may also contribute positively to innovative research initiatives such as Wolfram's search for a simple set of computer rules to explain how the universe works[59], or Davies' study of hidden webs of information in biological systems.

Finally, understanding the universe is more than a mere scientific curiosity. Seeing the bigger picture allows to put world views in perspective, provides greater clarity of mind and offers a solid foundation for efficient action. It seems appropriate, as a closing remark, to quote Einstein's thoughts on this issue:

> A human being is a spatially and temporally limited piece of the whole, what we call the "Universe". He experiences himself and his feelings as separate from the rest, an optical illusion of his consciousness. The quest for liberation from this bondage is the only object of true religion. Not nurturing the illusion but only overcoming it gives us the attainable measure of inner peace[60].

**Acknowledgements**

The author wishes to thank Timothée Samou for critical reading of the manuscript and René Oculi for production of the illustrations.